\documentstyle[12pt,aaspp4]{article}
\input psfig

\newcommand{\masyr}{mas$\,$yr$^{-1}$}
\newcommand{\fn}[1]{\tablenotemark{#1}}
\newcommand{\p}{\phm{mai)}}
\newcommand{\z}{\phm{mai}}
\newcommand{\y}{\phm{masi}}
\newcommand{\w}{\phm{ma}}

\newcommand{\n}{\phm{mi}}

\begin{document}
\def\i{\item}
\def\beq{\begin{equation}}
\def\eeq{\end{equation}}
\def\plotfiddle#1#2#3#4#5#6#7{\centering \leavevmode
\vbox to#2{\rule{0pt}{#2}}
\includegraphics{#1}}
 
\title{Proper motion of water masers associated with\\
IRAS 21391+5802: Bipolar Outflow\\
and an AU-scale Dusty Circumstellar Shell} 
\author{Nimesh A. Patel\altaffilmark{1}\\(npatel@cfa.harvard.edu)}
\author{Lincoln J. Greenhill\altaffilmark{1}\\(lgreenhill@cfa.harvard.edu)}
\author{James Herrnstein\altaffilmark{2}\\(jherrnstein@nrao.edu)}
\author{Qizhou Zhang\altaffilmark{1}\\(qzhang@cfa.harvard.edu)}
\author{James M. Moran\altaffilmark{1}\\(jmmoran@cfa.harvard.edu)}
\author{Paul T. P. Ho\altaffilmark{1}\\(pho@cfa.harvard.edu)}
\author{Paul F. Goldsmith\altaffilmark{3}\\(pfg@astrosun.tn.cornell.edu)}
\altaffiltext{1}{Harvard-Smithsonian Center for Astrophysics, 60 Garden St., Cambridge, MA 02138}
\altaffiltext{2}{National Radio Astronomy Observatory, P.O. Box 0, Socorro, NM
87801} 
\altaffiltext{3}{National Astronomy and Ionosphere Center, Cornell University, Department of Astronomy, Ithaca, NY 14853}

\begin{abstract}

We present VLBA observations
of water maser emission associated with the star forming region
IRAS 21391+5802, which is embedded in a bright rimmed cometary globule 
in IC1396. 
The angular resolution of the maps is $\sim$ 0.8 mas, 
corresponding to a spatial resolution of $\sim$0.6 AU, at an estimated
distance of 750 pc. Proper motions 
are derived for 10 maser features identified consistently over
three epochs, which were separated by intervals of about one month.
The masers appear in four groups, which are aligned 
linearly on the sky, roughly along a northeast--southwest direction,  
with a total separation of $\sim$520 AU ($\sim$0$\rlap{.}''$7). 
The 3D velocities of the masers have a maximum value of $\sim$42 km~s$^{-1}$ 
($\sim$9 AU yr$^{-1}$).
The average error on the derived proper motions is $\sim$4 km~s$^{-1}$.
The overall pattern of proper motions is indicative of a bipolar 
outflow. 
Proper motions of the
masers in a central cluster, with a projected extent of $\sim$ 20 AU, 
show systematic deviations from a radial outflow.
However, we find no evidence of Keplerian rotation, as has
been claimed elsewhere. A nearly circular loop of masers lies near the 
middle of the cluster.  The radius of this loop is 1 AU and the 
line-of-sight velocities of the masers in the loop are within 
2 km~s$^{-1}$ of the systemic velocity of the region.  
These masers presumably exist at the radial distance where
significant dust condensation occurs in the outflow emanating from
the star.  
\end{abstract}


\setcounter{footnote}{0}
\section{Introduction}
 
An important phase in the evolution of young stellar objects (YSO) 
is the one during which they exhibit energetic mass outflows, and 
simultaneous accretion of material onto the
star-forming core. Since the discovery of bipolar outflows (Snell et
al. 1980), much progress has been made in the study of the large scale 
and phenomenological characteristics of such outflows (e.g. Bachiller 1996),
but relatively fewer studies have  probed the regions within 
$\sim$ 100 AU of the YSOs,  where the winds that drive the outflows,
originate (e.g., Shu et al. 1994).

A significant fraction of all YSOs exhibit water maser emission,
and the maser luminosity appears to be correlated with the
mechanical luminosity of the outflows (Felli et al. 1992).
The maser emission is thought to arise from dust-laden shocked gas 
close to the exciting YSOs, where
densities are $\sim$ 10$^{9}$ cm$^{-3}$ and temperatures are $\sim$400 K 
(e.g. Elitzur et al. 1989). Since such conditions are likely to be
highly localized, and since the velocity coherence required for building
up large path lengths for the maser amplification is also likely to
occur over relatively smaller regions, the water maser emission provides
an excellent tool for probing the kinematics of the gas with high
angular resolution ($\sim$1 mas). Radio interferometric 
observations of the water maser emission 
and  measurements of maser proper motions have been carried out mostly towards 
high-mass star forming regions such as Orion-KL, W49N, W51 Main,  W51
North, Sgr B2 and W3~(OH) (Genzel et al. 1981a;  Gwinn et al. 1992; Genzel
et al. 1981b; Schneps et al. 1981; Reid et al. 1988; Alcolea et al.
1993).
Similar studies of low
and intermediate-mass ($<$10 M$_{\odot}$) star forming regions with
very high spatial resolution are only now becoming available 
(Claussen et al. 1998, Furuya et al. 1999, Wootten et al. 1999). 

IRAS 21391+5802 is an infrared source
embedded in a bright-rimmed globule appearing on the northern periphery
of the giant H {\sc ii} region IC1396.
Several observational studies of this object at various wavelengths 
suggest that it is an intermediate-mass 
star forming region (Sugitani et al. 1989, Wilking et al. 1993, 
Patel et al. 1995, Cesaroni et al. 1999). 
Its far-infrared luminosity is 235 L${_\odot}$ (Saraceno
et al. 1996).
Water maser emission from this source  was first detected by Felli et al.
(1992) and interferometric observations 
were made by Tofani et al. (1995) using the NRAO\footnote{The National Radio Astronomy Observatory is a facility
of the National Science Foundation operated under a cooperative agreement
with Associated Universities, Inc.}
Very Large Array (VLA)
at a resolution of 0$\rlap{.}''$1 resolution.
These observations
revealed a linear distribution
of masers, with a systematic velocity gradient along its length.
More recently, Slysh et al. (1999) reported single-epoch 
observations of these masers with mas angular resolution.
However, interpretation of the kinematics has remained ambiguous
without proper motion data.

We present here the results of a multi-epoch program of interferometric
observations of the water maser
emission from IRAS21391+5802.
In the next section, we summarize the observational details, followed
by results and our interpretation using a simple model of
a bipolar outflow. 
%Our conclusions are summarized in section 5.

%------------------------------end of introduction-------------------
\section{Observations and data reduction}

We observed the $6_{16}-5_{23}$
maser line at 22.23508 GHz, 
using the Haystack 37m telescope, the NRAO 140-foot antenna at 
Greenbank, WV, 
the VLA in its largest and smallest configurations,
and the NRAO Very Large Baseline Array (VLBA).
The single-dish observations in 1994-1995 provided a time-series of
spectra, while we determined absolute astrometric positions of the masers
using the VLA observations. 

The VLBA observations were made  for 6 hours on 
1996 March 17, 1996 April 20 and 1996 May 19.
We observed a single 4 MHz bandpass in left circular polarization,
  covering $-$27 to +27 km~s$^{-1}$ in velocity, 
and divided 
%which was
%spectroscopically analyzed 
into 256 channels of 0.21 km~s$^{-1}$ each.
The systemic velocity of the source, based on observations of 
$^{13}$CO J=1-0 emission,  with a velocity resolution of 0.7
km~s$^{-1}$, is about 0~km~s$^{-1}$ with respect to the Local Standard
of Rest (LSR) (Patel et al. 1995).
 The data were 
 correlated at the
NRAO Array Operation Center at Socorro, NM with an averaging time of
1.3 seconds, which provided a field-of-view of $\sim 1''$ for the whole
array.

The data were reduced and imaged with standard techniques using the
NRAO AIPS package.
Observations of 3C345, 1739+522, 2007+777, BL LAC, and 3C454.3
provided delay and phase calibration.
The sources 3C345 and BL Lac were used for bandpass 
calibration. 
%The AIPS tasks CALIB and IMAGR were used iteratively on this channel
%and the solutions were applied to all the channels. 
The synthesized beam in 
all three epochs 
was $\sim0.8\times 0.4$ mas at a position angle of $\sim20^{\circ}$.
The rms noise in the final images was $\sim$ 6 mJy in all 3
epochs.

Initially,  maps with reduced 1 $\times$ 1 mas resolution were made 
to identify the approximate locations of all the maser 
spots in a 1$'' \times 1''$ field. 
We define a maser ``spot'' as emission occurring in
a given velocity channel. A maser ``feature'' is a collection of spots
over contiguous channels.
Maps with the highest resolution were made for
selected regions and two dimensional elliptical Gaussians were fitted 
to each identified maser spot.
The positions of the masers were measured with respect to the spot at
$-$14.6 km~s$^{-1}$ in maser feature {\sc a}, which was used as 
a reference for self-calibration in all three epochs (i.e.,  for 
calibration of atmospheric path length fluctuations).  
The resulting position measurements have a mean formal uncertainty of
$\sim10 \mu$as.
For each spot, the adopted uncertainty is the larger of 1) the
theoretical uncertainty
from the beam-width and the signal-to-noise ratio,
and 2) the measured uncertainty from the profile fit.
Contributions to the position uncertainties from the
interferometric calibration (e.g., clock, astrometry, or baseline errors)
were negligible.
For each epoch, maser spot positions and line-of-sight velocities were 
inspected graphically to identify 
distinct features.
The number of identifiable maser features in the
three epochs was 21, 19,  and 20, respectively. 

Ten maser features persisted in all three epochs, with mean line-of-sight
velocities changing by less than 0.5 km~s$^{-1}$ ($\sim{1\over 2}$ linewidth).
Proper motions were estimated by a least-squares fit to position as a 
function of time, with weights that were inversely proportional to the
square of the uncertainties in position measurement. The weights were
scaled to achieve unit reduced $\chi^{2}$. A noise floor of 80 $\mu$as
was added in quadrature to the uncertainties to account for possible 
unresolved structure within the maser spots. The noise floor
corresponded approximately to the observed wander in the motion of the
maser features about a straight-line trajectory. Proper motions for
features observed at only two epochs were discarded because their
deviations from the straight-line motion could not be estimated.

%-----------------------end of observations and data reduction-----

\section{Results}

A time-series of spectra obtained during the time-monitoring observations 
of the water maser emission from IRAS 21391+5802 are shown in Fig. 1. 
The epochs are separated by $\ga$ 1 month.  
The spectral-lines change significantly from epoch to epoch. 
The center velocities of some lines
seem to change as well.
However, data obtained over shorter time intervals ($\le$ 2 weeks; not shown
here), do not show any significant changes.

The velocity structure of the spectra
are symmetric around the systemic velocity
of $\sim0$ km~s$^{-1}$, but the VLA images show that 
the 10 km~s$^{-1}$ emission is offset
southeast by $\sim$10$''$ (7500 AU) (see also
Tofani et al. 1995). Earlier interferometric observations of 
thermal continuum emission at 3mm, and C$^{18}$O J=1-0 line emission
made by Wilking et al. (1993), show
an extension from the central peak towards this 
$10$ km~s$^{-1}$ maser feature. We suggest that this
is a separate star forming condensation, 
unrelated to the rest of the masers in our map. 
Wilking et al. (1993) identified four near-infrared sources
towards IRAS21391+5802.  IRS2 (Wilking et al. 1993) which is 
associated with the 3 mm peak radio continuum emission,  is closest to
the water masers shown in  Fig. 2.
In the observations by Wilking et al. (1993), there does
not appear to be an infrared source associated with the $10$ km~s$^{-1}$
maser feature that is offset  10'' from the bulk of the maser
emission.

Fig. 2 shows the results of our VLBA observations over the three epochs. 
The groups of maser features labeled {\sc a}, {\sc c} and {\sc d} are isolated.
These masers have line-of-sight velocities differing by more than 
$\pm4$ km~s$^{-1}$ from the systemic velocity. 
{\sc a} and {\sc d} are the most extremely blue and red-shifted masers, respectively, 
and they also appear at the extrema of the distribution of masers on the sky.
Most of the masers in {\sc b} appear to be aligned along
an eastwest line near $\Delta\delta=0\rlap{.}''17$. We refer
to these hereafter as the ``inner" masers.
The integrated intensity maps of the inner
masers are shown in Fig. 3. From an analysis of fringe rates, we
estimate that the 
position of the reference emission is 
$\alpha_{J2000}=21^{h}40^{m}41^{s}.791\pm0^{s}.004,
\delta_{J2000}=58^{o}16'11\rlap{.}''737\pm0\rlap{.}''030$.
Our VLA A array observations made on 6 July 1995, add
information on the variability of maser emission in this region.
The maser features {\sc a} and {\sc b} appear roughly in the same
positions and line-of-sight velocities as in the VLBA observations.
Two different maser features appear in the VLA A array observations.
One at a position offset of $(0\rlap{.}''41,0\rlap{.}''11)$ with a
line-of-sight velocity of -8 km s$^{-1}$ and the other at
$(0\rlap{.}''63,0\rlap{.}''25)$ and velocity of 22 km s$^{-1}$.
Maser features {\sc c} and {\sc d} do not appear in the VLA map.
Because the position uncertainties are about $0\rlap{.}''1$ in
the VLA map, registered with respect to the VLBA map, the VLA data
are not useful in estimation of motions.

The proper motions in clumps {\sc a}, {\sc c} and {\sc d} are suggestive
of largely radial outflow from a center of expansion in clump {\sc b}.
Within clump {\sc b} there is a roughly circular loop of masers at 
$\Delta\alpha=0\rlap{.}''515-0\rlap{.}''519$ and 
$\Delta\delta=0\rlap{.}''16-0\rlap{.}''17$. The line-of-sight
velocities of these masers are the closest to the systematic velocity of
the region, and the velocities in the plane of the sky are consistent
with a radial outflow. 
From these results we propose that the loop signifies the radius of dust
sublimation in a circumstellar shell of outward flowing stellar wind material. 
Maser emission is beamed along the line-of-sight by the long gain paths 
tangent to the edge of the relatively dust-free inner cavity. 

From each feature, we have subtracted the mean 
measured proper motion for the whole source, to reduce the effect of 
the proper motion of the reference feature.
In principle, an arbitrary
velocity vector may be added to the proper motions since they are
obtained from positions to the feature at $-14.6$ km~s$^{-1}$. 
If we add a proper motion of 6.8 km~s$^{-1}$ at a
position angle of 29$^{\circ}$ to each maser feature, then the two proper
motion vectors associated with the loop become radial and outward.
However, with the addition of this largely northward motion, the proper
motions of the three easternmost features in {\sc b} deviate
significantly from a radial outflow.  Although the assignment of zero 
net proper motion may be somewhat arbitrary, it is the most straight-forward 
way to compensate at least partly for the relatively large apparent motion 
of the reference maser feature. 

%--------------------------------end of results------------------------

\section{Discussion}

The positions and 3D space velocities in Fig. 2 clearly show that
systematic motions dominate over turbulent motions in this system.
The masers trace bulk gas kinematics within 0$\rlap{.}''6$ or 500 AU
of the possible YSO.  
In principle, within such close proximity, both 
 infall along an accreting
circumstellar disk and an orthogonal  bipolar outflow 
may be anticipated  (e.g., Shu et al. 1987).
Relying on interpretation of spectra and a single 
epoch VLBA observation of these masers, Slysh et al.
(1999) conclude that the maser kinematics might represent
Keplerian rotation of a circumstellar disk around a YSO. 
However, in such a scenario the masers with line-of-sight velocity
nearest 
the systemic velocity should 
show the largest proper motions.
Our data clearly rule out a disk model  and support a
 bipolar outflow model.

The loop of masers  that we propose pinpoints the possible YSO (Figs. 2 \& 3) 
has a radius of $\sim$1 AU. 
The bolometric luminosity of IRAS 21391+5802 is 235 L$_{\odot}$,
implying a mass of 3 to 5 M$_{\odot}$. Gravitationally bound  gas within
a radius of a few AU of a 3 M$_{\odot}$ object should exhibit orbital
velocities of a few tens of km~s$^{-1}$. 
However, all the features in the loop have line-of-sight velocities that 
are an order of magnitude less. Furthermore, 
two of the features in the loop for which proper motions could be determined
show 3D space velocities of only $\sim$5 km~s$^{-1}$. 
If the masers lie at about the radius where
substantial condensation of dust grain occurs, then this is also
the radius at which acceleration of the outflow by radiation pressure
begins; the low 3D space velocities are reasonable and may be upper limits
on bulk outflow speeds closer to the YSO.  
Since the earliest observations of the maser source 
(Felli et al. 1992, Brand et al. 1994) the maser emission 
at line-of-sight velocities corresponding to the loop masers
has been persistent, in contrast to the higher velocity 
maser features, which apparently lie downstream in the
outflow.
The masers in
the loop may thus represent a standing pattern in outward flowing 
material, wherein masers are born, track outflowing and cooling
material, and fade, to be replaced by new masers at smaller radii.
Masers downstream are products of shocks and local heating possibly in
collisions between the outflow and ambient media.

The spectral energy distribution of the infrared emission from IRAS
21391+5802 has been studied by Correia et al. (1999). According
to their model, the temperature of the gas at a distance of 1 AU from
the YSO is predicted to be about 1800 K. This is in excess by at least a
factor of two higher than the temperature expected from the occurrence 
of the water masers in the loop. However, the calculated temperature depends
sensitively on grain properties such as size. If the grains are
larger than 1$\mu$m in size, the dust envelope would be optically
thin very close to the star, and lower temperatures would be possible
($\sim$1000 K).

The 3D space velocities of masers {\sc a}, {\sc c} and {\sc d} shown 
in Fig. 2 indicate an outflow away from the inner masers ({\sc b}).  
To estimate the  position and inclination angles of the outflow, we fit 
a simple model to the data in which we assume an outflow 
velocity that is linearly proportional to the distance from 
the outflow center, with a proportionality constant, $k$. 
We choose a coordinate system with its origin  at the center of the 
loop.
% the abscissa along the right ascension and the ordinate to be along 
%the declination, and the z axis is
%along  the line of sight. 
The purpose of this model is only to 
estimate the position and inclination angles and the value of $k$. 
A straight line fitted
to the observed maser positions alone provides
the position angle $\phi\approx 70^{\circ}$.
From the positions and 3D velocities, we obtain
an inclination angle $\theta\approx 70^{\circ}$ 
with respect to the line of sight, 
$k$, and $z_{i}$, the distance of each maser spot along
the line of sight, 
and $k\approx5\times 10^{-10} s^{-1}$ (0.07 km~s$^{-1}$ AU$^{-1}$).
The inclination and position angle are difficult to measure well because
the outflow appears to have an opening angle of at least 40$^{\circ}$,
judging from the motions and line-of-sight velocities of {\sc c} \& {\sc
D}, and from the distribution of maser spots in {\sc B}.

The fitted value of $k$ implies an overall e-folding  time of $\sim$65 yrs.
The dynamical times obtained from the de-projected radii of the masers
and the 3D space velocities are
43, 38 and 14 yrs for the masers {\sc a}, {\sc d} and {\sc c}. 
The inner masers ({\sc b}) have a dynamical time
of only 2 to 9 yrs. 
These different  dynamical times are inconsistent with the masers 
being ejected ballistically at the same time but with different initial 
velocities.
Instead, there is possibly true acceleration 
in the flow as a function of distance away from the
YSO. Another possibility is that the flow is ballistic but episodic.
However, as noted earlier, maser emission at velocities close
to the systemic velocity (corresponding to {\sc b} \& {\sc l}) seems to
have persisted since the earliest observations in 1988 
(Brand et al. 1994). Thus, the mass
outflow represented by the clumps of gas that are masing today, may be a
continuous process, rather than a consequence of a singular event of 
mass ejection.

%-----------------------------end of discussion-----------------------

\section{Conclusions}

We have detected proper motion of 10 water maser features associated 
with the star-forming region IRAS 21391+5802. 
The water masers trace a bipolar outflow
from an intermediate mass  young stellar object. 
%The masers have a mean 
%magnitude of 3D space velocity of $\sim$13 km~s$^{-1}$ (2.7 AU yr$^{-1}$). 
We estimate position and inclination angles of $\sim 70^{\circ}$ for the
outflow axis, though the opening angle may exceed 40$^{\circ}$.
The maximum observed 3D space velocity of the masers is 42 km~s$^{-1}$. 
The observed 3D space velocities of the masers suggest the presence of 
acceleration within the outflow, but we cannot rule out episodic ballistic
ejection.
Near the center of the flow, there is a roughly circular loop of masers 
with a radius of $\sim$1 AU. 
The masers in this loop are most likely to be
tangentially amplified within a shell of dense gas
surrounding the YSO. Because there is no evidence for rotation in the
source (as in a disk), the shell probably comprises wind material
alone.
However, dynamics of the IRAS 21391+5802 region are poorly sampled,
especially in the loop. Additional 
observations may still reveal some rotation, or even infall.

\acknowledgments
We thank Lee Hartmann, Nuria Calvet, T. K. Sridharan and Masao Saito for 
helpful discussions.
We are grateful to Shoshana Rosenthal for help with the management of the
Alpha workstations and AIPS in the Radio and Geoastronomy division 
at the Center for Astrophysics.

%--------------------------------end of conclusion------------------

\newpage
\clearpage

\begin{deluxetable}{rcccrrrrc}
\tablecaption{Positions, velocities and proper motions of masers}
\tablewidth{0pt}
\tablehead{
\colhead{} & \multicolumn{2}{c}{Position offset\fn{a}} && \multicolumn{4}{c}{Proper motion\fn{b}} & Maser\fn{c}\\
\cline{2-3} \cline{5-9} \\
\colhead{V$_{LSR}$} & \colhead{$\Delta \alpha$} & \colhead{$\Delta \delta$} &&
                 \colhead{$\dot{\alpha}$} & \colhead{Error} &
                 \colhead{$\dot{\delta}$} & \colhead{Error}& \\
\colhead{(km$\,$s$^{-1}$)} & \colhead{(mas)} & \colhead{(mas)} &&
		 \colhead{(\masyr)} & \colhead{(\masyr)} & \colhead{(\masyr)} & \colhead{(\masyr)} &}
\startdata
-14.6 \n & 0.9 & 0.2 && -10.2\z & 1.2 \p & -2.5 \y & 0.2\w & A\\
-10.0 \n & 529.4 & 191.6 &&&&&&B\\
-9.4 \n & 527.0 & 225.2 && 3.2\z & 3.8 \p & 2.4 \y & 0.2\w & C\\
-9.1 \n & 529.0 & 190.1 &&&&&&B\\
-8.6 \n & 529.0 & 190.1 &&&&&&B\\
-8.0 \n & 528.4 & 190.0 &&&&&&B\\
-7.6 \n & 527.7 & 189.7 &&&&&&B\\
-7.6 \n & 529.3 & 190.0 &&&&&&B\\
-3.7 \n & 536.9 & 183.7 &&&&&&B\\
-1.9 \n & 535.0 & 182.3 &&&&&&B\\
-1.4 \n & 506.8 & 167.6 && -1.8\z & 0.6 \p & -1.7 \y & 0.02\w & B\\
-1.0 \n & 509.1 & 168.6 && -0.3\z & 0.5 \p & -2.2 \y & 0.1\w & B\\
-0.9 \n & 528.0 & 178.5 &&&&&&B\\
-0.5 \n & 515.3 & 168.1 &&&&&&B\\
-0.5 \n & 506.9 & 167.9 &&&&&&L\\
-0.0 \n & 516.6 & 168.0 &&&&&&L\\
0.0 \n & 516.4 & 167.8 &&&&&&L\\
0.2 \n & 516.5 & 167.8 &&&&&&L\\
0.3 \n & 516.7 & 167.8 &&&&&&L\\
0.4 \n & 515.5 & 169.3 && 0.3\z & 0.3 \p & -1.3 \y & 0.4\w & L\\
0.4 \n & 517.8 & 168.9 &&&&&&L\\
0.4 \n & 516.9 & 168.0 &&&&&&B\\
0.4 \n & 507.0 & 168.0 &&&&&&B\\
0.8 \n & 522.4 & 162.7 &&&&&&L\\
0.8 \n & 516.2 & 167.0 &&&&&&L\\
\tablebreak
0.9 \n & 517.1 & 166.9 &&&&&&L\\
1.0 \n & 517.1 & 169.8 && 1.5\z & 0.7 \p & -0.8 \y & 0.2\w & L\\
1.0 \n & 517.6 & 167.9 &&&&&&L\\
1.1 \n & 516.5 & 166.9 &&&&&&L\\
1.1 \n & 517.4 & 167.5 &&&&&&L\\
1.4 \n & 527.0 & 168.8 && 0.2\z & 1.0 \p & 1.7 \y & 0.1\w & B\\
1.5 \n & 526.3 & 161.2 &&&&&&B\\
2.0 \n & 528.5 & 175.8 &&&&&&B\\
2.1 \n & 531.6 & 169.6 && 1.6\z & 0.2 \p & 0.7 \y & 0.2\w & B\\
2.5 \n & 523.0 & 168.7 && 1.0\z & 0.4 \p & 0.8 \y & 0.4\w & B\\
7.5 \n & 615.9 & 334.9 && 4.5\z & 0.3 \p & 3.0 \y & 1.2\w & D\\
\enddata
\tablenotetext{a}{Position in right ascension and declination offset
from the reference feature (V$_{LSR}$=-14.6 km~s$^{-1}$), the estimated
position of which is 
$\alpha_{J2000}=21^{h}40^{m}41^{s}.791\pm0^{s}.004,
\delta_{J2000}=58^{o}16'11\rlap{.}''737\pm0\rlap{.}''030$}
\tablenotetext{b}{A proper motion vector of 38.7 km~s$^{-1}$ at a
position angle of 256$^{\circ}$ has been added to the gross motions to
achieve a zero mean (see text).}
\tablenotetext{c}{See labels in Figure 2}
\end{deluxetable}
\newpage

% FIGURE 1
\begin{figure}[t]
\plotfiddle{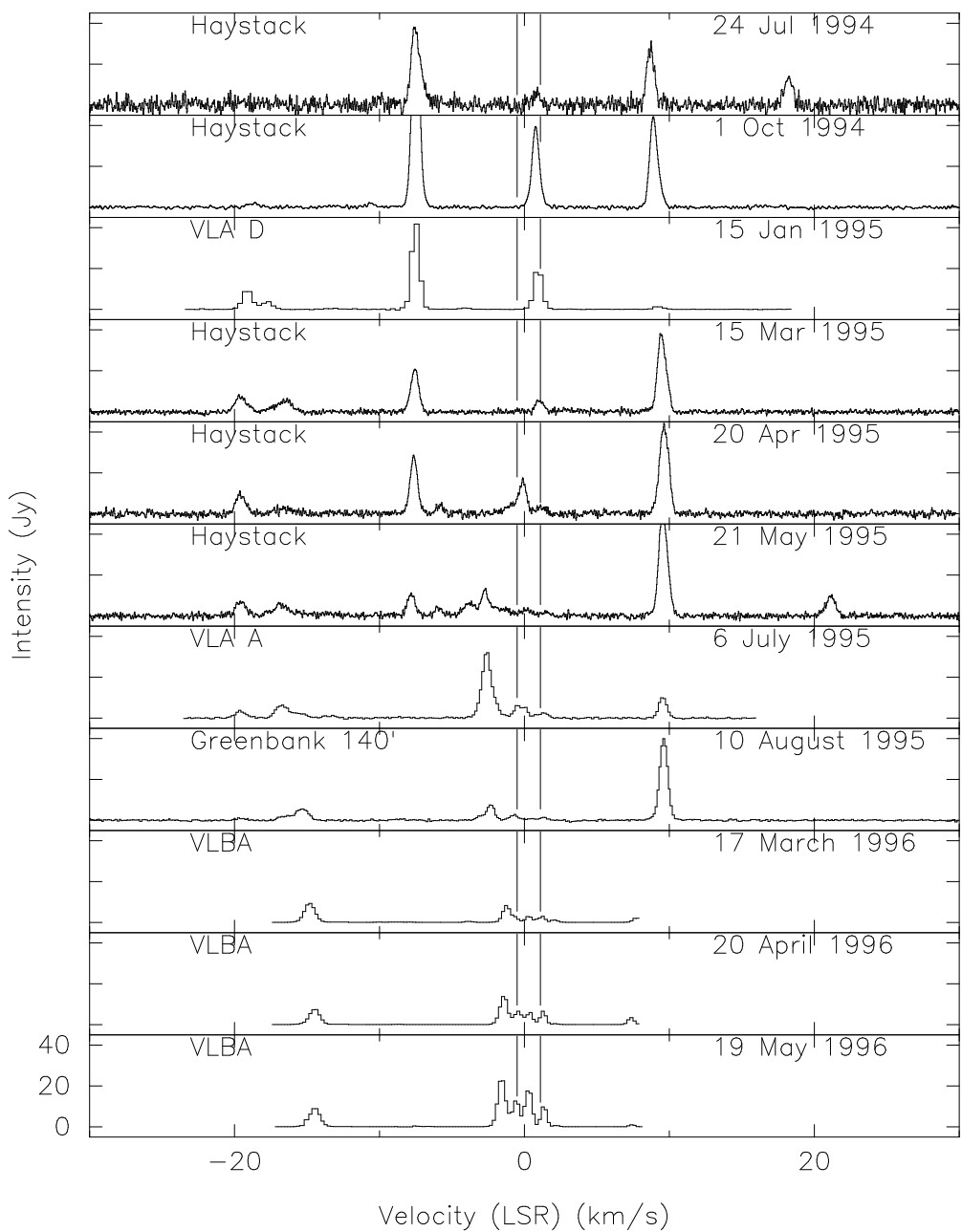}{6.75in}{0}{70}{70}{-220}{-50}
\caption{\setlength{\baselineskip}{8 pt} \footnotesize
Total power spectra obtained at the Green Bank 140-foot and Haystack 37m
antennas and integrated VLA and VLBA spectra of water maser 
emission associated with IRAS 21391+5802.
The maser features vary significantly in intensity on time scales
greater than a month. Line-of-sight velocities may also change on
time scales of months. The VLBA spectra covered the velocity range of
$-$27 to 27 km~s$^{-1}$ but no emission was found at velocities less than
$-$14 km~s$^{-1}$ and greater than 8 km~s$^{-1}$. (The VLBA data are
 shown only for the range of spectral
 channels that display maser emission on the dates indicated).
The vertical lines indicate the range of line-of-sight velocities of the
masers occurring in the loop shown in Fig. 2. The systemic velocity of
the source is $\sim$ 0 km~s$^{-1}$.
}
\end{figure}
\newpage

% FIGURE 2
\begin{figure}[t]
\plotfiddle{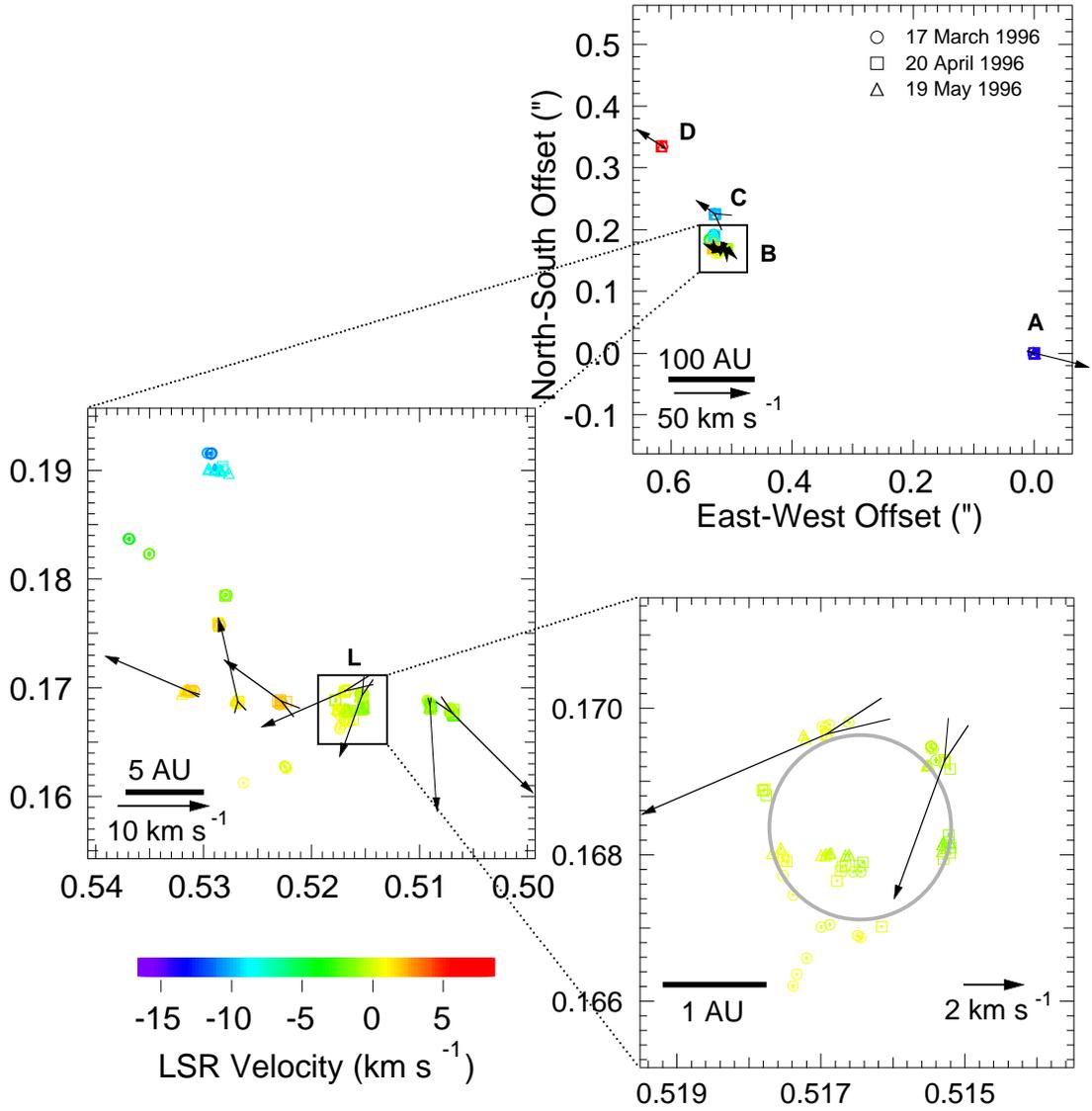}{5.9in}{0}{75}{75}{-225}{-140}
\caption{\setlength{\baselineskip}{8 pt} \footnotesize
Maps of fitted water maser spots and maser feature proper motions 
from all 3 VLBA epochs superposed. {\it Upper right:} 
Positions and proper motions of the water masers
are largely indicative of a bipolar outflow motion.
The formal and systematic errors in these positions are
much smaller than the size of the symbols in the plot. 
The $1\sigma$ uncertainties in the direction of proper motion 
are indicated by the angle
indicated at the tails of the arrows. The length of the cones
imply $1\sigma$ uncertainties in the magnitude of the proper motion.
Line-of-sight velocities are indicated by symbol  colors, according to
the scale bar  at the bottom left.
The maser labeled {\sc a} is the reference feature in all three epochs.
{\it Middle left:} The proper motions and positions of the inner masers ({\sc b}), are shown 
on an enlarged scale in the lower panel. 
{\it Lower right:} The masers roughly in the center of the
{\sc b} trace a nearly circular loop; the YSO is probably located at the
center. The radius of the loop of masers is $\sim$1 AU. 
The loop most likely represents the inner
radius of dust condensation in a shell of outflowing material around the YSO.
}
\end{figure}
\newpage

% FIGURE 3
\begin{figure}[t]
\plotfiddle{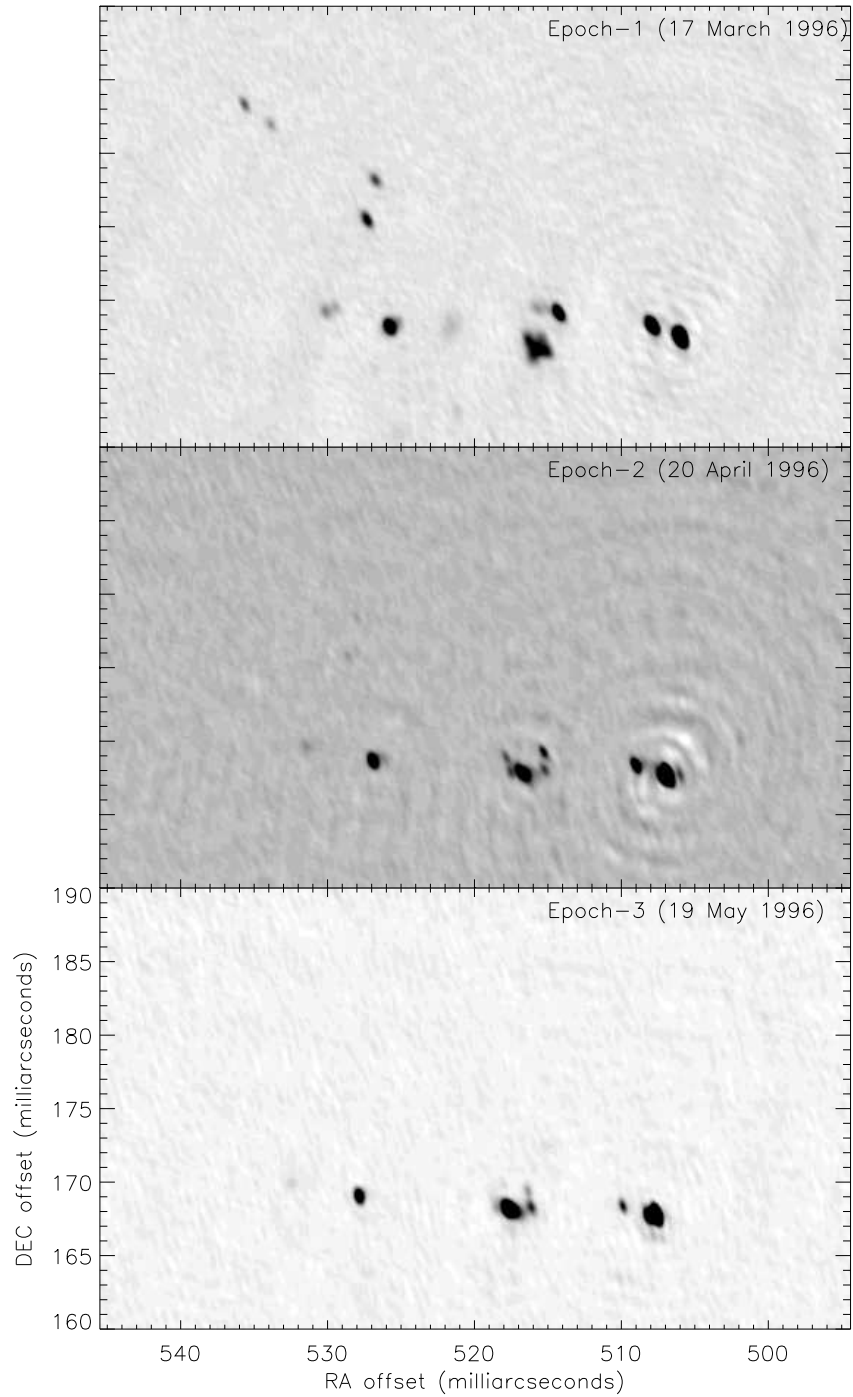}{7.25in}{0}{80}{80}{-265}{-40}
\caption{\setlength{\baselineskip}{8 pt} \footnotesize
Integrated intensity images at three epochs, for emission  between
V$_{LSR}=-$4.6 and 1.7 km~s$^{-1}$ within  the central clump of masers.
The overall variability  of features in the spectra (Fig. 1) is consistent with
changes among the maps.
The masers near the position offset $\Delta\alpha=0\rlap{.}''516,
\Delta\delta=0\rlap{.}''168$ correspond to the loop pictured in Fig. 2.
}
\end{figure}
\newpage

% FIGURE 4
\begin{figure}[t]
\plotfiddle{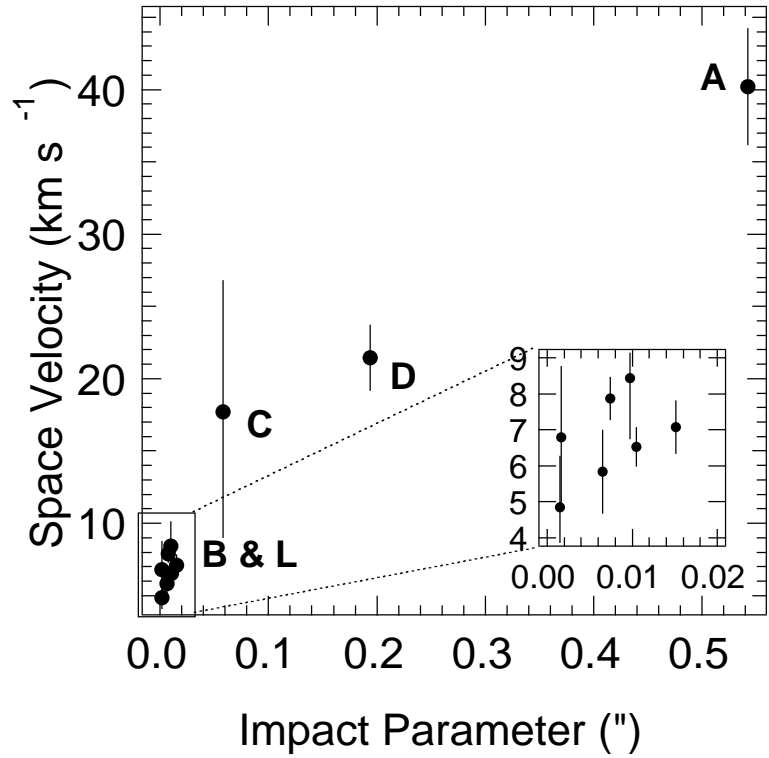}{5in}{0}{100}{100}{-310}{0}
\caption{\setlength{\baselineskip}{8 pt} \footnotesize
Variation of 3D space velocity as a function of distance from the 
position of the YSO
(which is assumed to be at the center of the loop of masers shown in Fig. 2).
This figure shows that 1) the masers that are more distant from the YSO,
move relatively faster than the closer ones and 2) the dynamical time for
the inner masers ({\sc b} \& {\sc l}) is less than 
that for the outer masers ({\sc a}, {\sc c} \& {\sc d}).
}
\end{figure}

\end{document}